\documentclass[aps,reprint,superscriptaddress,a4paper,showpacs,showkeys,prl]{revtex4-1}
\usepackage{graphicx} 
\usepackage{dcolumn} 
\usepackage{bm}
\usepackage{hyperref} 
\usepackage{color} 
\usepackage[dvipsnames]{xcolor}
\usepackage{amsmath}
\usepackage{amssymb} 
\usepackage[normalem]{ulem}

\begin{document}

\title{Apparent Contact Angle and Contact Angle Hysteresis on Liquid Infused Surfaces}

\author{Ciro Semprebon} \email{ciro.semprebon@durham.ac.uk} 
\affiliation{Department of Physics, Durham University, Durham, DH1 3LE, UK}

\author{Glen McHale} 
\affiliation{Smart Materials \& Surfaces Laboratory, Faculty of Engineering \& Environment, Northumbria University, Newcastle upon Tyne NE1 8ST, UK}

\author{Halim Kusumaatmaja} \email{halim.kusumaatmaja@durham.ac.uk} 
\affiliation{Department of Physics, Durham University, Durham, DH1 3LE, UK}

\begin{abstract}
We theoretically investigate the apparent contact angle and 
contact angle hysteresis of a droplet placed on a liquid infused surface.
We show that the apparent contact angle is not uniquely defined by material parameters, but also has a dependence on 
the relative size between the droplet and its surrounding wetting ridge formed by the infusing liquid. 
We derive a closed form expression for the contact angle in the limit of vanishing wetting ridge, and compute the
correction for small but finite ridge, which corresponds to an effective line tension term. We also predict 
contact angle hysteresis on liquid infused surfaces generated by the pinning of the contact lines by the 
surface corrugations. Our analytical expressions for both the apparent contact angle and contact angle hysteresis 
can be interpreted as `weighted sums' between the contact angles of the infusing liquid relative to the 
droplet and surrounding gas phases, where the weighting coefficients are given by ratios of the fluid surface tensions. 
\end{abstract}

\maketitle

\section{Introduction}
\label{sec:Intro}

\begin{figure}[tbh]
\centering
\includegraphics[width=1.0\columnwidth]{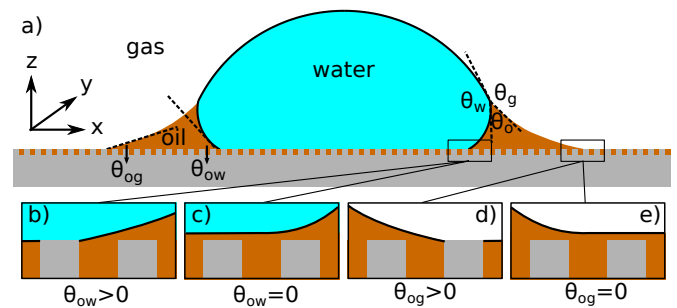}
\caption{(a) The geometry of an axisymmetric water droplet placed on a liquid infused surface, indicating
the three Neumann angles $\theta_{\rm o}$, $\theta_{\rm w}$ and $\theta_{\rm g}$,
and the wetting angles relative to the oil phase $\theta_{\rm ow}$ and $\theta_{\rm og}$.
Depending on the oil-water contact angle, the water droplet can touch the solid substrate (b) 
or be separated by a thin oil layer (c). Similarly, the oil-gas contact angle determines whether
the infusing oil merely hemi-wicks the rough surface (d) or coat the surface corrugations (e).
Based on the combinations of (b-c) and (d-e), four wetting states of interest on liquid infused 
surfaces can be distinguished.
}
\label{fgr:sketch_LIS}
\end{figure}

A novel class of functional surfaces has recently been introduced, known as Liquid Infused Surfaces\cite{Epstein2012} (LIS), 
Lubricant Impregnated surfaces\cite{Smith2013} (LIS), or Slippery Liquid Infused Porous Surfaces\cite{Wong2011} (SLIPS).
They are typically constructed by infusing rough or porous materials with lyophilic oils.
Thus, as shown in Fig. \ref{fgr:sketch_LIS}, a typical liquid infused system involves three fluids: 
the oil phase, and typically a (water) droplet in a gas environment.
They have been shown to exhibit many advantageous wetting properties, including low contact angle 
hysteresis, self-cleaning, drag reduction, anti-icing and anti-fouling
\cite{Wong2011,Lafuma2011,Kim2012,Epstein2012,Smith2013,Schellenberger2015,Wexler2015}. 
Furthermore, compared to competing technologies such as superhydrophobic surfaces, 
LIS are robust against pressure-induced instabilities and failure
\cite{Kusumaatmaja2008a,Hejazi2011,Butt2013,Tuteja2008a}, which makes them
favourable for applications to a wide-range of problems, ranging from marine fouling and product 
packaging to heat exchanger and medical devices\cite{Leslie2014,Xiao2013,Epstein2012}.

For LIS, oil penetration in between the surface corrugations is key. As such, the material contact 
angles characterizing the infusing oil relative to the air and water phases have to satisfy two wicking criteria: 
$\theta_{\rm ow}^{\rm Y}<\theta^\ast$ for the oil--water and
$\theta_{\rm og}^{\rm Y}<\theta^\ast$ for the oil--gas interfaces.
In general the critical angle $\theta^\ast$ will depend on the details of the surface corrugations\cite{Semprebon2014a}, 
but in many cases of interest a simple thermodynamic criterion can be expressed in terms of global statistical quantities. 
These are the roughness factor $r$, corresponding to the ratio of total area of the textured surface to its projected area, 
and the fraction $f$ of the projected area that is occupied by a solid\cite{Bico2002,Quere2008}
\begin{equation} \label{eq:wicking}
\cos \theta^{\ast} = \frac{(1-f)}{(r-f)}.
\end{equation}
It is worth noting that, even with these thermodynamic criteria satisfied, impalement of the water and gas phases
may still occur at high pressures\cite{Kusumaatmaja2008a,Hejazi2011,Butt2013}, but we will exclude such cases here.

For vanishing contact angles $\theta^{\rm Y}_{\rm ow} = 0^\circ$ or $\theta^{\rm Y}_{\rm og} = 0^\circ$, 
it is possible that a thin oil layer will completely cover the surface roughness.
Provided the wicking criteria are satisfied, we can identify four relevant thermodynamic 
wetting states\cite{Smith2013}. As illustrated in Fig. \ref{fgr:sketch_LIS}(b-e)
these states depend on the presence and absence of a thin oil film between the water and/or gas phases and the solid surface.
The presence of a thin oil film prevents a direct contact between the water/gas phase
and the solid. It has been proposed to explain the smooth displacement of contact lines on 
liquid infused surfaces\cite{Guan2015}, as opposed to stick-slip commonly observed on other surfaces\cite{Varagnolo2013}.
 
While thermodynamics arguments are sufficient to predict the presence of different wetting states on LIS, 
to date there is no theory for computing the corresponding values of the contact angle 
and contact angle hysteresis, despite their relevance as key design parameters for any 
application. For example, low contact angle and low contact angle hysteresis are preferred 
for efficient heat transfer \cite{Xiao2013,Rykaczewski2014}, while high contact angle and low contact 
angle hysteresis are desirable for high droplet mobility\cite{Mahadevan1999}.

For standard wetting scenarios, the Young's equation determines how the contact angle depends 
on the three independent (solid-liquid, solid-gas and liquid-gas) surface tensions. In contrast, 
there are six independent surface tensions for LIS, and an equivalent relation for the contact 
angle as a function of these surface tensions is, to date, not yet available. We will derive 
such a relation in this paper. Furthermore, an oil ridge is drawn to the 
oil-water-gas three-phase contact line in LIS system due to capillary action, as depicted in Fig. \ref{fgr:sketch_LIS}(a). 
It is unclear how the shape and size of this ridge can be controlled, and correspondingly 
what the consequences might be. Indeed, a distinguishing feature of LIS we will show here is that 
both the apparent contact angle and contact angle hysteresis are not uniquely defined by material parameters; 
instead, they also have a strong dependence on the size of the oil ridge relative to the droplet, 
which in return can be manipulated by tuning the oil pressure.

This paper is organised as follows. In section 2, we will list the essential
physical assumptions for the theoretical model and describe the computational method we employ 
to calculate drop morphologies on liquid infused surfaces.
We will derive a closed form expression for the contact angle in the limit 
of vanishing oil ridge in section 3. In section 4, supported by numerical
calculations, we will address the influence of oil pressure on the apparent contact angle.
In section 5, we will discuss how the theoretical results in sections 3 and 4
can be extended to predict contact angle hysteresis generated by contact line pinning
at the edges of the surface corrugations.
Finally, we will conclude and describe future works in section 6.

\section{Physical model and numerical method}
\label{sec:Model}

\subsection{Physical model}

For concreteness, let us consider a typical LIS system consisting of a water droplet (w) 
deposited on a porous/rough solid substrate (s) infused by an oil liquid (o), and immersed 
in a surrounding gas phase (g) as shown in Fig. \ref{fgr:sketch_LIS}. Our theory 
is valid, without any loss of generality, if other fluids are used instead of water, oil and gas. 
Let us also define $\gamma_{\rm wg}$, $\gamma_{\rm ow}$ and $\gamma_{\rm og}$ as the surface 
tensions between the water--gas, oil--water and oil--gas components respectively. 
We further assume that the typical length scales in the problem (the droplet size, the oil 
ridge, and the surface corrugation) are smaller than the capillary length, such that 
we can neglect gravity. For water and most oils, the capillary length is of the order of a few millimetres.

The total energy $E_{\rm b}$ has several different contributions. First, this energy contains 
two terms that depend on the volumes of the water droplet $V_{\rm w}$ and infusing oil $V_{\rm o}$, 
and on the pressure differences $\Delta P_{\rm wg}$ between the water droplet and 
the surrounding gas, and $\Delta P_{\rm og}$ between the oil and the surrounding gas.
It is also convenient to define $\Delta P_{\rm ow}$ as the 
pressure difference between oil and water. 
Second, each fluid-fluid interface contributes with a term proportional to 
$\gamma_{\alpha\beta}\,A_{\alpha\beta}$. 
The subscripts $\alpha,\beta$ correspond to the water, oil and gas phases. 
Third, if any of the phases is in contact with a portion of the solid substrate, it also 
contributes with a term $\gamma_{\alpha \rm s}\,A_{\alpha \rm s}$, where ${\rm s}$ indicates 
the solid surface. Thus, the total energy is given by: 
\begin{equation}
E_{\rm b} = \Delta P_{\rm wg} V_{\rm w} + \Delta P_{\rm og} V_{\rm o} + \sum_{\alpha\neq \beta} \gamma_{\alpha\beta} A_{\alpha\beta} + \sum_{\alpha} \gamma_{\alpha \rm s} A_{\alpha \rm s}. 
\label{eq:energy}
\end{equation}
Let us now discuss the suitable ensemble for the water and oil phases. 
Usually the volume of the water droplet is fixed in experiments.
As such, in our calculations the pressure difference $\Delta P_{\rm wg}$ acts as a Lagrange multiplier 
to the droplet volume.
For the oil phase, instead it is appropriate to assume the pressure ensemble due to 
the presence of a large amount of oil infused in between the surface 
corrugations, which to a good approximation can be considered as an infinite reservoir.
In this context, the definition of $V_{\rm o}$ in Eq. \eqref{eq:energy} corresponds to the
amount of oil drawn from the reservoir into the ridge. The oil which fills the surface
roughness is not included in the computation of $V_{\rm o}$.

In equilibrium, the Laplace pressures of the fluid-fluid interfaces determine their mean curvatures 
$\kappa$ through the Laplace law 
\begin{equation}
\Delta P_{\rm \alpha\beta}=2\kappa_{\rm \alpha\beta}\gamma_{\rm \alpha\beta}.
\label{eq:Laplace}
\end{equation}
As before, the subscripts $\alpha,\beta$ correspond to the water, oil and gas phases. 
At the triple point junction, where the three fluid interfaces meet, the stresses are balanced
\begin{equation}
\vec{\gamma}_{\rm ow}+\vec{\gamma}_{\rm og}+\vec{\gamma}_{\rm wg} = 0.
\label{eq:balance}
\end{equation}
As shown in Fig. \ref{fgr:sketch_LIS}(a), Eq. \eqref{eq:balance} 
leads to the Neumann angles\cite{Rowlinson}, $\theta_{\rm o}$, $\theta_{\rm w}$ and $\theta_{\rm g}$, where
\begin{equation}
\frac{\gamma_{\rm ow}}{\sin \theta_{\rm g}} = \frac{\gamma_{\rm wg}}{\sin \theta_{\rm o}} 
= \frac{\gamma_{\rm og}}{\sin \theta_{\rm w}},
\label{eq:Neumann}
\end{equation}
and $\theta_o+\theta_w+\theta_g=2\pi$.
It is worth noting that, for $\gamma_{\rm wg} > \gamma_{\rm ow} + \gamma_{\rm og}$, the water 
droplet is encapsulated by a thin layer of oil, and Eq. \eqref{eq:Neumann} is ill-defined. 
In this work, we will exclude such a case, and assume that the Neumann angles can be computed 
according to Eq. \eqref{eq:Neumann} for a given set of water--gas, oil--gas, and water--oil surface tensions.

The interaction between the ternary fluids (water--oil--gas) with the (smooth) solid surface 
can be characterised by three material contact angles, $\theta^{\rm Y}_{\rm wg}$, $\theta^{\rm Y}_{\rm ow}$ 
and $\theta^{\rm Y}_{\rm og}$, given by the Young's relation
\begin{equation}
\cos{\theta^{\rm Y}_{\rm \alpha\beta}} = \frac{\gamma_{\rm \beta s}-\gamma_{\rm \alpha s}}{\gamma_{\rm \alpha\beta}}, 
\end{equation}
where once again the subscripts $\alpha,\beta$ correspond to the water, oil and gas phases. 
For liquid infused surfaces, the solid surface is not smooth. In fact, the surface roughness is 
key for maintaining the infusing oil. As shown in Fig. \ref{fgr:sketch_LIS}, a typical 
substrate can be modelled as a composite between solid and oil. 
If pinning effects and the related energy barriers are negligible, the contact 
angles can be described by weighted averages as proposed 
by Cassie and Baxter \cite{Cassie1944}, 
\begin{equation} 
\cos{\theta^{\rm CB}_{\alpha\beta}} = f \cos{\theta^{\rm Y}_{\rm \alpha\beta}} +(1-f), 
\label{eq:CB} 
\end{equation}
where $\alpha$ represents the oil phase, $\beta$ is either the water or gas phase,
and $f$ is the fraction of the projected solid area exposed to the water or gas phase.
We will explicitly consider pinning phenomena at the sharp edges of the surface roughness
in section \ref{sec:CAH}, where we address the emergence of contact angle hysteresis on LIS.
Formally the weighted average will enter the total energy $E_{\rm b}$ by redefining the
surface energy of the composite substrate
$\gamma_{\alpha \rm s} \rightarrow f  \gamma_{\alpha \rm s} +(1-f)$,
where $\alpha$ represents the water or gas phase.
For the Cassie-Baxter equation to be valid, the water droplet needs to cover 
a sufficiently large number (e.g. several tens) of posts. 

As it is not the aim of this work to resolve the liquid morphology down to the molecular scale, 
we will not specifically model a thin oil film on top of the roughness, because the effect of
a thin microscopic film on the equilibrium shape of a macroscopic droplet is negligible.
Its presence, however,  will affect the choice of $\theta_{\rm ow}^{\rm Y}$ and 
$\theta_{\rm og}^{\rm Y}$. Typically it is convenient to assume the value of 
$0^\circ$ when a thin wetting oil film is present, though in general a thin oil film does not 
necessarily lead to vanishing contact angles, depending on the effective interfacial potentials 
describing the intermolecular forces acting between the substrate and the water and oil 
molecules \cite{Dufour2016,Kuchin2014}. 

\subsection{Numerical method}

In our calculations we assume the distortion induced by the underlying pattern geometry to be 
negligible and the drop to retain an axial symmetry. We will numerically compute droplet 
configurations by employing a finite element approach based on the free software SURFACE EVOLVER \cite{Brakke1996}. 
Eq. \eqref{eq:energy} will be minimized according to standard minimisation algorithms.
Without losing generality, we will set the reference pressure of the gas phase to zero,
while the pressures of the water and oil phases will correspond to the Laplace pressures
of the water--gas and oil--gas interfaces. The oil phase will be controlled by its
pressure $\Delta P_{\rm og}$, by including the corresponding term  in the total energy in 
Eq. \eqref{eq:energy}, while the volume $V_{\rm w}$ of the water droplet will be imposed by a 
global constraint. The Laplace Pressure $\Delta P_{\rm wg}$ is the Lagrange multiplier to
the constant droplet volume constraint.

\section{The limit of vanishing oil ridge}
\label{sec:2Dsolution}

In the most general case, the shapes of fluid-fluid interfaces with constant mean curvature 
and an axial symmetry must belong to the family of Delaunay surfaces\cite{Delaunay1841}. 
In our problem, the water-gas interface will be a portion of sphere, while the oil-water and oil-gas interfaces can 
be described either by nodoids or unduloids, depending on the boundary conditions (wetting contact angles). 
In this section we are interested in the case where the size of the oil ridge is infinitely small compared to the water
drop. For most liquid infused surfaces, this limit is equivalent to the condition of large and negative oil 
pressure in comparison to the Laplace pressure in the water droplet, $-\Delta P_{\rm wg}/\Delta P_{\rm og}\rightarrow 0$,
since $\Delta P_{\rm og} < 0$ is the regime of physical interest.
In section 4, we will comment on the implications related to the case of $\Delta P_{\rm og} > 0$.
In the $-\Delta P_{\rm wg}/\Delta P_{\rm og}\rightarrow 0$ limit,
the geometry near the oil ridge can be simplified. The water-gas interface is
effectively flat. The curvature in the $x-y$ plane for the oil-water and oil-gas interfaces can be neglected.
Their profiles in the $x-z$ plane are circular arcs to an excellent approximation.

\begin{figure}[tb]
\centering
  \includegraphics[width=0.85\columnwidth]{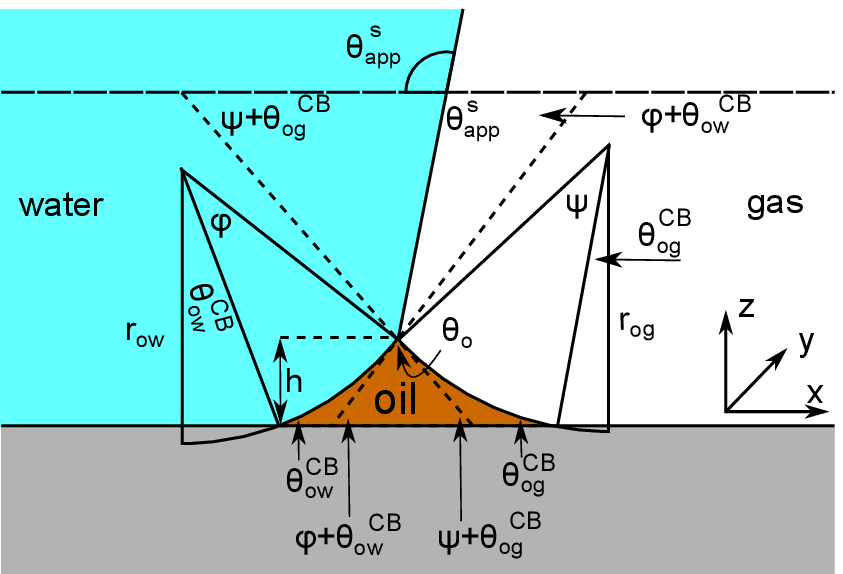}
  \caption{Sketch illustrating the derivation of the closed form expression for $\theta^S_{\rm app}$. In the limit of small
  oil ridge the water-gas interface is flat in proximity of the Neumann triangle, while the oil-water and oil-gas interfaces
  can be approximated by circular arcs. The surface corrugation is not explicitly depicted in this sketch. However, its effect
  is accounted by using the Cassie-Baxter's rather than the Young's angles for the oil-water and oi-gas interfaces.
  }
  \label{fgr:sketch2D}
\end{figure}

Referring to the sketch shown in Fig. \ref{fgr:sketch2D}, we introduce two auxiliary angles $\varphi$ and $\psi$
for the water-oil and oil-gas interfaces. These interfaces respectively have radii of curvature $r_{\rm ow}$ and $r_{\rm og}$.
The oil-water and oil-gas  interfaces approach the substrate with contact angles $\theta^{\rm CB}_{\rm ow}$ and 
$\theta^{\rm CB}_{\rm og}$. Since $-\Delta P_{\rm wg}/\Delta P_{\rm og}\rightarrow 0$, we can deduce that 
\begin{equation}
-\Delta P_{\rm ow} = \frac{\gamma_{\rm ow}}{r_{\rm ow}} =  -\Delta P_{\rm og} = \frac{\gamma_{\rm og}}{r_{\rm og}},
\label{eq:radiusratio0}
\end{equation}
which, combined with Eq. \eqref{eq:Neumann}, leads to
\begin{equation}
\frac{r_{\rm ow}}{r_{\rm og}} = \frac{\gamma_{\rm ow}}{\gamma_{\rm og}}=\frac{\sin\theta_{\rm  g}}{\sin\theta_{\rm  w}}.
\label{eq:radiusratio}
\end{equation}
In Fig. \ref{fgr:sketch2D}, we can identify a triangle with interior angles given by 
$\varphi+\theta^{\rm CB}_{\rm ow}$, $\psi+\theta^{\rm CB}_{\rm og}$ and $\theta_{\rm o}$.
Imposing their sum to be $\pi$, we have a trigonometrical relation
\begin{equation}
\label{eq:phipsirelation}
\varphi+\theta^{\rm CB}_{\rm ow}+\psi+\theta^{\rm CB}_{\rm og}+\theta_{\rm o}=\pi.
\end{equation}
Similarly we can derive geometrical relations for the apparent contact angle $\theta^S_{\rm app}$:
\begin{equation}
\label{eq:relationrecedingpressure}
\theta^S_{\rm app}=\theta_{\rm w}-\theta^{\rm CB}_{\rm ow}-\varphi=\pi-\theta_{\rm g}+\theta^{\rm CB}_{\rm og}+\psi.
\end{equation}
Here we have used the superscript $S$ to denote the limiting case of vanishing oil volume.
To complete the set of equations and express $\theta^S_{\rm app}$ in terms of the remaining material parameters, 
we can deduce one more equation by comparing the expressions for the height $h$ given by both menisci,
\begin{eqnarray}
h &= & r_{\rm og}\left[\cos\theta^{\rm CB}_{\rm og}-\cos(\psi+\theta^{\rm CB}_{\rm og})\right] \nonumber \\
  &= & r_{\rm ow}\left[\cos\theta^{\rm CB}_{\rm ow}-\cos(\varphi+\theta^{\rm CB}_{\rm ow})\right].
\label{eq:hconditionrec}
\end{eqnarray}
Substituting Eqs. \eqref{eq:radiusratio}, \eqref{eq:phipsirelation} and \eqref{eq:relationrecedingpressure} 
into Eq. \eqref{eq:hconditionrec}, we obtain
\begin{eqnarray}
\label{eq:anglepressureeq}
&\sin\theta_{\rm w} \left[\cos\theta^{\rm CB}_{\rm og}+\cos(\theta^S_{\rm app}+\theta_{\rm g})\right]= \nonumber \\
&\sin\theta_{\rm g} \left[\cos\theta^{\rm CB}_{\rm ow}-\cos(\theta_{\rm w}-\theta^S_{\rm app})\right].
\end{eqnarray}
Eq. \eqref{eq:anglepressureeq} can be inverted to express $\theta^S_{\rm app}$ only in terms of the 
Neumann, oil-water and oil-gas contact angles:
\begin{equation}
\label{eq:anglepressure}
\cos\theta^S_{\rm app} =\left(\frac{\cos\theta^{\rm CB}_{\rm ow}\sin\theta_{\rm g} - \cos\theta^{\rm CB}_{\rm og} \sin\theta_{\rm w}}
{\cos\theta_{\rm g}\sin\theta_{\rm w} + \cos\theta_{\rm w} \sin\theta_{\rm g}}\right).
\end{equation}
Note that the specific value of the Laplace pressures for the oil-water and oil-gas interface
do not appear in Eq. \eqref{eq:anglepressure}. 

It is convenient to express the apparent contact angle in Eq. \eqref{eq:anglepressure} in terms of the
fluid-fluid surface tensions and the oil-water and oil-gas contact angles. This is the form 
in which the material properties are most commonly reported and tabulated. 
To do this, we observe that the denominator in Eq. \eqref{eq:anglepressure} can be simplified as
$\sin (\theta_{\rm g} + \theta_{\rm w}) = \sin (2\pi - \theta_{\rm o}) = -\sin (\theta_{\rm o})$. 
Taking further advantage of Eq. \eqref{eq:Neumann}, we obtain
\begin{equation}
\cos\theta^S_{\rm app} = - \cos\theta^{\rm CB}_{\rm ow} \frac{\gamma_{\rm ow}}{\gamma_{\rm wg}} + \cos\theta^{\rm CB}_{\rm og} \frac{\gamma_{\rm og}}{\gamma_{\rm wg}},
\label{eq:angle2}
\end{equation}
or alternatively
\begin{equation}
\cos\theta^S_{\rm app} = \cos\theta^{\rm CB}_{\rm wo} \frac{\gamma_{\rm ow}}{\gamma_{\rm wg}} + \cos\theta^{\rm CB}_{\rm og} \frac{\gamma_{\rm og}}{\gamma_{\rm wg}},
\label{eq:angle3}
\end{equation}
where we have used $\cos\theta^{\rm CB}_{\rm wo} =-\cos\theta^{\rm CB}_{\rm ow} $.

In general Eq. \eqref{eq:angle2} can be interpreted as a `weighted sum' between the oil-water and oil-gas 
contact angles, where the weighting coefficients are given by ratios of the fluid-fluid surface tensions. 
Several limiting cases are worth being pointed out:
(i) When the oil-gas surface tension is very small, such that $\gamma_{\rm og} \rightarrow 0$ and
$\gamma_{\rm ow} \rightarrow \gamma_{\rm wg}$, we recover 
$\cos\theta^S_{\rm app} = - \cos\theta^{\rm CB}_{\rm ow} =  \cos\theta^{\rm CB}_{\rm wo}$. 
This is the Cassie-Baxter contact angle for a water droplet on a composite solid-oil substrate;
(ii) Similarly, in the limit of $\gamma_{\rm ow} \rightarrow 0$ and $\gamma_{\rm og} \rightarrow \gamma_{\rm wg}$, 
we recover the Cassie-Baxter angle, $\cos\theta^S_{\rm app} = \cos\theta^{\rm CB}_{\rm og}$;
(iii) For $\theta^{\rm CB}_{\rm ow}\rightarrow 0$ and  $\theta^{\rm CB}_{\rm og}\rightarrow 0$,
we recover a condition equivalent to the Young's equation, 
$\cos\theta^S_{\rm app} = (\gamma_{\rm og} - \gamma_{\rm ow})/\gamma_{\rm wg}$, corresponding to a water 
droplet spreading on a flat substrate made of oil;
(iv) The apparent contact angle approaches $\theta^S_{\rm app} \sim 90^\circ$ if the oil-gas and oil-water 
interfaces have `symmetric' properties,
$\gamma_{\rm og} \sim \gamma_{\rm ow}$ and $\theta^{\rm CB}_{\rm ow} \sim \theta^{\rm CB}_{\rm og}$, 
irrespective of their actual values.

\section{Role of the Laplace pressures}
\label{sec:Angles}

\begin{figure}[tb]
\centering
  \includegraphics[width=0.6\columnwidth]{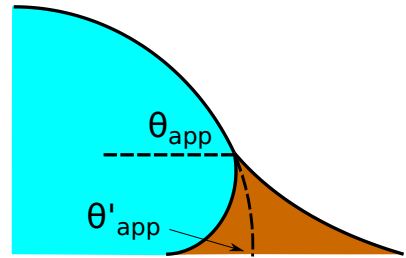}
  \caption{Sketch of a drop with finite oil ridge displaying two possible definitions 
  for the apparent angle: $\theta_{\rm app}$ at the triple junction of
  the fluid phases and $\theta'_{\rm app}$ at the virtual contact line where the 
  interpolated water-gas interface meets the solid substrate.}
  \label{fgr:definition_angles}
\end{figure}

In the previous section we derived an analytical expression for the apparent contact angle $\theta_{\rm app}^{\rm S}$
in the limit of small oil ridge, assuming a vanishing Laplace pressure for the water-gas interface. 
In this section we will extend our analysis to the general case, explicitly accounting for the role of finite Laplace pressures 
for the three interfaces. In particular we will show that $\theta_{\rm app}$ is not uniquely determined by the material parameters,
but it is affected by the relative size of the oil ridge relative to the size of the water drop. 

\begin{figure}[tb]
\centering
  \includegraphics[width=0.8\columnwidth]{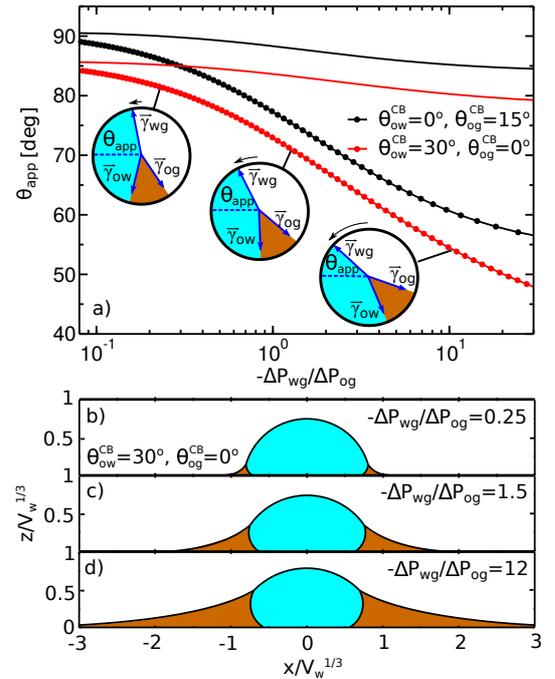}
  \caption{Numerical results for a water droplet placed on liquid infused surfaces.
  The Neumann angles for the oil, water and gas phases are respectively
  $\theta_{\rm o}=30^\circ$ and $\theta_{\rm w}=\theta_{\rm g}=165^\circ$. 
  Two sets of wetting angles are considered: 
  (i) $\theta^{\rm CB}_{\rm ow}=0^\circ$ and $\theta^{\rm CB}_{\rm og}=15^\circ$, and   
  (ii) $\theta^{\rm CB}_{\rm ow}=30^\circ$ and $\theta^{\rm CB}_{\rm og}=0^\circ$.   
  a) Variation of the apparent contact angle as a function of $-\Delta P_{\rm wg}/\Delta P_{\rm og}$.
  The results for both definitions of apparent contact angles are shown,
  $\theta_{\rm app}$ (points) and $\theta'_{\rm app}$ (straight lines).
  The insets illustrate how the Neumann triangles rotate as the oil pressure is varied.
  b-d) Snapshots of numerically evaluated ternary drop configurations. 
   }
  \label{fgr:pressure}
\end{figure}

To illustrate the role of the Laplace pressures,  we have numerically 
computed ternary drop morphologies for representative systems typical of an oil with low 
surface tension, choosing $\theta_{\rm o}=30^\circ$ and symmetrically 
$\theta_{\rm w}=\theta_{\rm g}=165^\circ$ (see Fig. \ref{fgr:pressure}). 
We consider two different combinations of wetting angles for the oil phase, given by 
(i) $\theta^{\rm CB}_{\rm ow}=0^\circ$, $\theta^{\rm CB}_{\rm og}=15^\circ$ and 
(ii) $\theta^{\rm CB}_{\rm ow}=30^\circ$, $\theta^{\rm CB}_{\rm og}=0^\circ$.
We then vary the oil pressure while keeping the water volume constant.
In our calculations the specific value of the volume is not relevant, as it 
simply sets the length scale of the system.

In the presence of finite Laplace pressures it is necessary to adapt the definition
of $\theta_{\rm app}$ as the water-gas interface is no longer represented by a straight line. 
Two meaningful geometric choices are possible as shown in Fig. \ref{fgr:definition_angles},
either (i) as the slope of the water-gas interface at the triple junction $\theta_{\rm app}$, or (ii)
as the slope of the virtual water-gas interface as it is extrapolated down to the
solid substrate $\theta'_{\rm app}$. The water-gas interface assumes a spherical
cap geometry and the extrapolation procedure is unique.
We find $\theta'_{\rm app}>\theta_{\rm app}$.
In the limit of vanishing oil ridge, both definitions converge to the same value
corresponding to Eq. \ref{eq:angle3}, which describes the energy balance at the contact line.
The deviation grows larger with increasing size of the ridge.

In this work we favour the apparent contact angle definition at the triple junction, 
$\theta_{\rm app}$, for two main reasons:
(a) the angle at the Neumann triangle can be easily identified from the kink in the drop profile, 
and it can be directly measured both in simulations and experiments, and 
(b) it represents a direct measure of the rigid rotation of the Neumann triangle with respect to 
the solid surface. In contrast, $\theta'_{\rm app}$ describes the slope of a portion of an interface 
that is only virtual, and cannot be measured directly.

The sequence of morphologies reported in Fig. \ref{fgr:pressure}(b-d) shows 
the impact of increasing $-\Delta P_{\rm wg}/\Delta P_{\rm og}$ to the 
growth of the oil ridge. As depicted in Fig. \ref{fgr:pressure}(a), for the chosen
sets of parameters, this is accompanied with a decrease in $\theta_{\rm app}$, 
as consequence of the rigid rotation of the Neumann triangle (see the insets). 
For the two specific examples we have shown here, the variation in the apparent contact angle
between $-\Delta P_{\rm wg}/\Delta P_{\rm og}\rightarrow 0$ (small ridge) and 
$-\Delta P_{\rm wg}/\Delta P_{\rm og}\rightarrow  \infty$ (large ridge) is above $30^\circ$. 
Similarly $\theta'_{\rm app}$ also decreases with increasing $-\Delta P_{\rm wg}/\Delta P_{\rm og}$
but its variation is considerably smaller, limited to a few degrees. 
It is worth noting that our definition for the apparent contact angle is intended to characterise not just the water droplet shape,
but instead the combined water droplet-oil ridge configuration, which spreads out as $-\Delta P_{\rm wg}/\Delta P_{\rm og}$ increases. In this context, $\theta_{\rm app}$ captures this behaviour better than $\theta'_{\rm app}$.
From here on, we will only focus on $\theta_{\rm app}$.

One aspect differentiating the two combinations of angles reported in Fig. \ref{fgr:pressure} is worth 
discussing further. In both cases the
limit $-\Delta P_{\rm wg}/\Delta P_{\rm og}\rightarrow  \infty$ (i.e. $\Delta P_{\rm og}\rightarrow 0$) 
implies that the oil-gas ridge approaches
a catenoid shape. However, while for finite contact angle $\theta^{\rm CB}_{\rm og}$
the oil ridge has a finite size, in the case of $\theta^{\rm CB}_{\rm og}=0^\circ$ the
radius of the oil ridge is diverging. The latter corresponds to a liquid lens configuration,
where the Neumann triangle is oriented such that the oil-gas interface is flat and lies parallel to the
solid substrate.


For small but finite $-\Delta P_{\rm wg}/\Delta P_{\rm og}$, we are also able to derive a closed form
expression for the apparent contact angle $\theta_{\rm app}$. 
To proceed, we note that the profiles of the oil-gas and 
oil-water interfaces can be still be assumed to be circular arcs in the $x-z$ plane, as shown in 
Fig. \ref{fgr:sketch2D}. Following the convention displayed in Fig. 2, we have $r_{\rm og} > 0$ and 
$r_{\rm ow} > 0$ when $\Delta P_{\rm og} < 0$ and $\Delta P_{\rm ow} < 0$.
The curvature in the $x-y$ plane can also be neglected for the oil-gas and oil-water interfaces.
Taking into account the Laplace pressure difference (or the curvature) of the water-gas interface, 
we can write
\begin{equation}
\Delta P_{\rm wg}=\frac{\gamma_{\rm ow}}{r_{\rm ow}}-\frac{\gamma_{\rm og}}{r_{\rm og}}.
\label{eq:Laplace_balance}
\end{equation}
A straightforward manipulation invoking Eq. \eqref{eq:Laplace} leads to the following equation,
\begin{equation}
\frac{r_{\rm og}}{r_{\rm ow}}=\frac{\gamma_{\rm og}}{\gamma_{\rm ow}} \left(1-  \frac{\Delta P_{\rm wg}}{\Delta P_{\rm og}}  \right).
\label{eq:pressure_parameter}
\end{equation}
\vspace{0.0cm}

\begin{figure}[tb]
\centering
  \includegraphics[width=0.8\columnwidth]{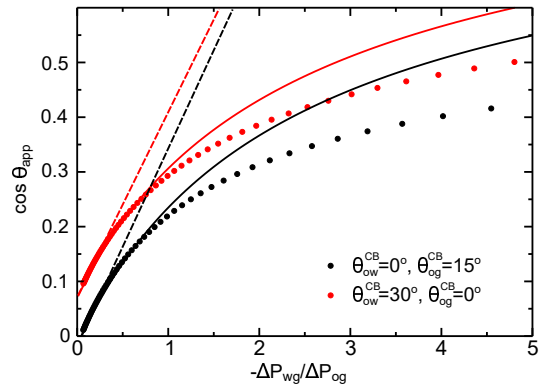}
  \caption{Comparison between numerical results (dots), Eq. \eqref{eq:anglepresssol0} 
  (full lines), and Eq. \eqref{eq:linetension} (dashed lines). The analytical expressions
  are valid for small $-\Delta P_{\rm wg}/\Delta P_{\rm og}$.
  The system parameters are the same as those reported in Fig. \ref{fgr:pressure}. 
    }
  \label{fgr:compare_evolver_model}
\end{figure}

Following the same route leading to Eq. \eqref{eq:anglepressureeq} in the previous section,
we can write down an equivalent relation with a correction term due to finite $-\Delta P_{\rm wg}/\Delta P_{\rm og}$,
\begin{equation}
\frac{\sin\theta_{\rm g} \left[\cos\theta^{\rm CB}_{\rm ow}-\cos(\theta_{\rm w}-\theta_{\rm app})\right]}
{\sin\theta_{\rm w} \left[\cos\theta^{\rm CB}_{\rm og}+\cos(\theta_{\rm app}+\theta_{\rm g})\right]}
=\left(1-  \frac{\Delta P_{\rm wg}}{\Delta P_{\rm og}}  \right).
\label{eq:anglepressureeq2}
\end{equation}
This relation can be inverted for the apparent contact angle, and it is given by
\begin{equation}
\cos\theta_{\rm app}=\frac{A C + B\sqrt{A^2 + B^2 - C^2}}{A^2 + B^2},
\label{eq:anglepresssol0}
\end{equation}
where
\begin{eqnarray}
&A=\sin\theta_{\rm g} \cos\theta_{\rm w}  + \left(1-  \frac{\Delta P_{\rm wg}}{\Delta P_{\rm og}}  \right) \sin\theta_{\rm w}  \cos\theta_{\rm g}, \label{eq:anglepresssol1} \\
&B=\sin\theta_{\rm g} \sin\theta_{\rm w} \left(\frac{\Delta P_{\rm wg}}{\Delta P_{\rm og}}  \right), \label{eq:anglepresssol2} \\
&C=\sin\theta_{\rm g}  \cos\theta^{\rm CB}_{\rm ow}  - \left(1-  \frac{\Delta P_{\rm wg}}{\Delta P_{\rm og}}  \right) \sin\theta_{\rm w}  \cos\theta^{\rm CB}_{\rm og} . \label{eq:anglepresssol3}
\end{eqnarray}
As we can observe in Fig. \ref{fgr:compare_evolver_model}, the analytical expression compares well with the full numerical results
for $ -\Delta P_{\rm wg}/\Delta P_{\rm og}<1$. For larger $ -\Delta P_{\rm wg}/\Delta P_{\rm og}$, the model departs from the numerical solution, 
as the circular arc approximation for the oil-water and oil-gas interfaces breaks down.

Furthermore, it is useful to extract a linear correction to the apparent contact angle due to the parameter $-\Delta P_{\rm wg}/\Delta P_{\rm og}$,
given by
\begin{widetext}
\begin{equation}
\cos\theta_{\rm app}=\cos\theta_{\rm app}^{\rm S} + \Lambda \left( \frac{\Delta P_{\rm wg}} {\Delta P_{\rm og}}  \right) 
+ \mathcal{O}\left( \frac{\Delta P_{\rm wg}} {\Delta P_{\rm og}}  \right)^2,
\label{eq:anglepressexpansion}
\end{equation}
with
\begin{equation}
\Lambda= \frac{\sin\theta_{\rm g}\sin\theta_{\rm w} \left(\cos\theta_{\rm g} \cos\theta^{\rm CB}_{\rm ow} + \cos\theta^{\rm CB}_{\rm og} \cos\theta_{\rm w} + 
   \sqrt{\sin(\theta_{\rm g} + \theta_{\rm w})^2 -(\cos\theta^{\rm CB}_{\rm ow} \sin\theta_{\rm g} - \cos\theta^{\rm CB}_{\rm og} \sin\theta_{\rm w})^2 }\right)}
   {\sin(\theta_{\rm g}+\theta_{\rm w})^2}.
\label{eq:chipar}
\end{equation}
\end{widetext}
In experiments, the pressure difference between the oil and gas phases is usually kept constant. The Laplace pressure
of the water droplet is given by $\Delta P_{\rm wg} = 2 \gamma_{\rm wg}/r_{\rm wg} \simeq  2\gamma_{\rm wg}\sin\theta^S_{\rm app}/R$,
where $R/r_{\rm wg} \simeq  \sin\theta^S_{\rm app}$ is the effective contact radius, taken at the Neumann's triple junction between the
water, oil and gas phases. Since we are only interested in the first order correction here, we have also approximated $\theta_{\rm app}\simeq\theta^S_{\rm app}$.
As such, Eq. \eqref{eq:anglepressexpansion} can be written as
\begin{equation}
\cos\theta_{\rm app}=\cos\theta_{\rm app}^{\rm S} + \frac{2\Lambda\gamma_{\rm wg}\sin\theta^S_{\rm app}}{\Delta P_{\rm og}}  \times \frac{1}{R},
\label{eq:linetension}
\end{equation}
This equation has a familiar interpretation in the literature of wetting phenomena: 
it is reminiscent to the correction term in Young's angle due to line tension \cite{Gaydos1987,Marmur1997}, 
with an effective line tension given by
\begin{equation}
\tau =  -\frac{2\Lambda\gamma^2_{\rm wg}\sin\theta^{\rm S}_{\rm app}}{\Delta P_{\rm og}}.
\label{eq:linetension2}
\end{equation}
The linear approximation in Eq. \eqref{eq:linetension} is also shown in Fig. \ref{fgr:compare_evolver_model}. It is in good agreement with the
full numerical results for $ -\Delta P_{\rm wg}/\Delta P_{\rm og}<0.5$. 

\begin{figure}[tb]
\centering
  \includegraphics[width=0.8\columnwidth]{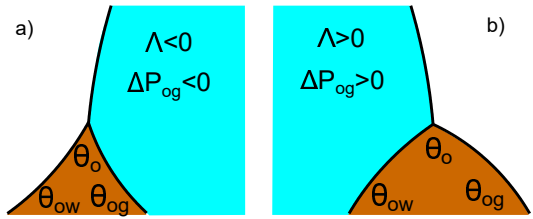}
  \caption{Sketch illustrating the stable shapes of the oil ridge depending on the sign of $\Lambda$:
  a) the common case, with $\Lambda<0$ and $\Delta P_{\rm og}<0$; and
  b) with $\Lambda>0$ and $\Delta P_{\rm og}>0$
    }
  \label{fgr:press_pos}
\end{figure}

The coefficient $\Lambda$ in Eq. \eqref{eq:chipar} can in principle assume both positive and negative values.
It can be shown that $\Lambda<0$ if $\theta_{\rm w}+\theta_{\rm g}>\pi +\theta_{\rm ow} +\theta_{\rm og}$. 
Using the fact that $\theta_{\rm w}+\theta_{\rm g} +\theta_{\rm o}=2\pi$,
we obtain $\pi > \theta_{\rm o} +\theta_{\rm ow} +\theta_{\rm og}$, which has a simple geometrical interpretation.
From Fig. \ref{fgr:press_pos} it is clear that when $\theta_{\rm o} +\theta_{\rm ow} +\theta_{\rm og}=\pi$
the three angles of the oil ridge form a triangle.
When $\Lambda<0$ the sum of these angles is smaller than $\pi$, and the oil ridge is stable only if 
$\Delta P_{\rm og}<0$ and $\Delta P_{\rm ow}<0$.
In contrast, if $\Lambda>0$, the sum of the three angles is larger than $\pi$, and the ridge is stable only if 
$\Delta P_{\rm og}>0$ and $\Delta P_{\rm ow}>0$. 
$\Lambda<0$ and $\Delta P_{\rm og}<0$ represent the most relevant physical regime for liquid infused surfaces.
The case of $\Lambda>0$ implies larger oil-water and oil-gas wetting angles, which are often in conflict with the 
wicking criterion in Eq. \eqref{eq:wicking}. Additionally, for $\Delta P_{\rm og}>0$, the fluid configuration 
could be unstable against non-axisymmetric perturbations\cite{Schafle2010}.

\begin{figure}[tb]
\centering
  \includegraphics[width=0.95\columnwidth]{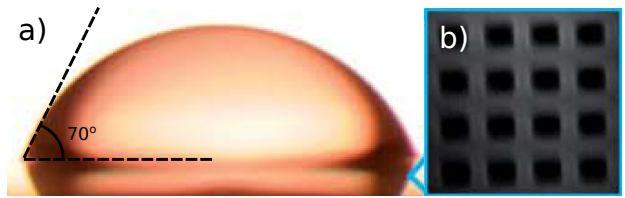}
  \caption{a) Experimental image of a water droplet on an OTS surface infused by BMIm as the lubricant, taken from Smith et al. \cite{Smith2013}. 
  The observed apparent contact angle $\theta_{\rm app}\simeq 70^\circ \pm 2^\circ$ is in excellent agreement with the prediction 
  of Eq. \ref{eq:angle2} $\theta_{\rm app}^{\rm S}= 70.9^\circ$. b) The surface pattern used in the experiment, again taken from Smith et al. \cite{Smith2013}. 
  From panel b), we estimate that the projected solid fraction exposed to the water and gas phases is $f=0.44$.
       }
  \label{fgr:exp_angle}
\end{figure}

Taking advantage of Eqs. \eqref{eq:angle2} and \eqref{eq:linetension2}, we have computed the apparent 
contact angles and effective line tensions for several LIS systems reported in the literature\cite{Wong2011,Anand2015,Smith2013} 
in Table \ref{tbl:example}. In Fig. \ref{fgr:exp_angle}, we have also shown one experimental drop morphology from Smith et al. \cite{Smith2013}. Here the oil ridge is small in comparison to the droplet size, and as such, we expect Eq. \ref{eq:angle2} to provide an excellent approximation. Indeed the measured contact angle is in agreement with the theoretical prediction, assuming $f=0.44$.

To compute the effective line tension in Table \ref{tbl:example}, we have assumed a typical Laplace pressure 
$|\Delta P_{\rm og}|=10^3$ Pa for the oil--gas ridge, corresponding to a radius of curvature of 
$r_{\rm og} \sim 100 \, \rm{\mu m}$. The computed effective line tension values are comparable to 
those measured for gas-liquid-solid contact line tensions\cite{Amirfazli2004}. Noticeably,
liquid infused surfaces always have negative effective line tensions since the signs of $\Lambda$ and
$\Delta P_{\rm og}$ are always the same for the system to be stable.
Thus, the apparent contact angle of a water droplet on a liquid infused surface increases with increasing droplet volume.
Our analytical expressions are readily applicable to other solid surfaces and fluids 
(both for the droplet and the lubricant) for cases where $\gamma_{\rm wg} < \gamma_{\rm ow} + \gamma_{\rm og}$.

\begin{table*}
\small
  \begin{tabular*}{\textwidth}{@{\extracolsep{\fill}}ccccccccccccccc}
    \hline
    Source & Solid & droplet (w) & lubricant (o) & $\gamma_{\rm wg}$ & $\gamma_{\rm og}$ & $\gamma_{\rm wo}$  
    & $\theta_{\rm w}$  & $\theta_{\rm o}$  & $\theta_{\rm g}$  & $\theta_{\rm ow}$  & $\theta_{\rm og} $  
     & $\theta_{\rm app}^{\rm S}$  & $\Lambda$  & $\tau$ \\
    \hline
    Ref. \cite{Schellenberger2015} & inv. opal  & H$_2$O& decanol  & $30$ & $28.5$ & $8.6$  & $108.3$ & $88.4$ & $163.3$ & $0$ & $0$ & $48.4$ & $-0.14$ & $-1.9\times 10^{-7}$ \\
    Ref. \cite{Smith2013} & OTS  & H$_2$O& BMIm  & $42$ & $34$ & $13$  & $135.4$ & $60.2$ & $164.4$ & $37$ & $64$ & $70.9$ & $-0.15$ & $-4.9\times 10^{-7}$ \\
    Ref. \cite{Wong2011} & S.Epoxy & H$_2$O & FC-70 & $72.4$ & $17.1$ & $56.0$  & $175.6$ & $18.8$ & $165.6$ & $36.5$ & $14.1$ & $118.2$ & $-0.29$ & $-2.7\times 10^{-6}$ \\
    Ref. \cite{Wong2011} & Epoxy & H$_2$O & FC-70 & $72.4$ & $17.1$ & $56.0$  & $175.6$ & $18.8$ & $165.6$ & $71.7$ & $33.5$ & $108.7$ & $-0.24$ & $-2.3\times 10^{-6}$ \\
    Ref. \cite{Wong2011} & Silicon & C16 H$_34$  & H$_2$O & $27.2$ & $72.4$ & $51.1$  & $47.2$ & $164.0$ & $48.8$ & $5.6$ & $13.1$ & $28.8$ & $0.028$ & $-2.0\times 10^{-8}$ \\
    \hline
  \end{tabular*}
  \caption{Theoretical prediction for the apparent contact angle $\theta_{\rm app}^{\rm S}$ for 
  several LIS systems reported in the literature\cite{Wong2011,Anand2015,Smith2013},
  as given by Eq. \eqref{eq:angle2}. For the Cassie-Baxter contact angles, we have assumed rough surfaces with projected solid area fraction $f=0.44$. 
  The surface tensions are expressed in the unit of mN/m, the line tension $\tau$ is in Newton (N), and the angles are in degrees ($^\circ$). 
  $\Lambda$ is the dimensionless parameter needed for computing the line tension as defined in Eq. \eqref{eq:chipar}.
  For the computation of the line tension in Eq. \eqref{eq:linetension2}, we have assumed
  a typical Laplace pressure $\Delta P_{\rm og}=10^3$ Pa for the oil--gas ridge, corresponding to a radius of curvature, $r_{\rm og} \sim 100 \, \rm{\mu m}$. 
  }
  \label{tbl:example}
\end{table*}

\section{Contact Angle Hysteresis}
\label{sec:CAH}

In the previous sections we computed the apparent contact angles
in thermodynamic equilibrium, by introducing the Cassie-Baxter contact angles on the composite substrate.
In this section we will address how pinning of the oil-water and oil-gas contact lines give rise to contact angle
hysteresis on liquid infused surfaces. In general contact line pinning can be generated either by chemical
heterogeneities or surface topographies\cite{Kusumaatmaja2007,Semprebon2012,Lenz1998,Oliver1977,Johnson1964}. 
Here we will focus on the latter. Following the Gibbs condition \cite{Gibbs}, a pinned contact line does not exhibit a unique contact angle, 
instead it can take a range of values.

\begin{figure}[tb]
\centering
  \includegraphics[width=0.99\columnwidth]{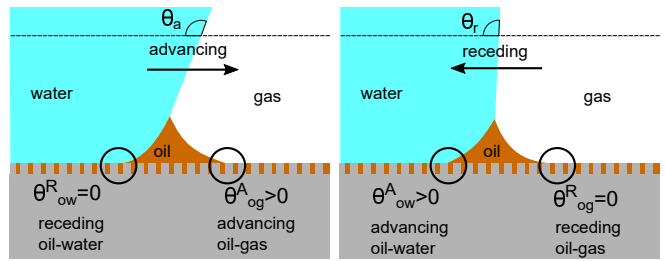}
  \caption{Contact angle hysteresis on liquid infused surfaces arises due to the pinning of oil-water 
  and oil-gas contact lines by the surface corrugations. 
  Sketches illustrating the configurations of the oil ridge for (a) an advancing and (b) a receding water droplet.
 }
  \label{fgr:sketch_hysteresis}
\end{figure}

There are four wetting states on liquid infused surfaces:
(i) When $\theta^{\rm CB}_{\rm ow} = \theta^{\rm CB}_{\rm og} = 0^\circ$, we expect the contact angle hysteresis to be negligible;
(ii) In contrast, for $\theta^{\rm CB}_{\rm ow} > 0^\circ$ and $\theta^{\rm CB}_{\rm og} > 0^\circ$,
the oil-water and oil-gas contact lines can both be pinned. As illustrated in Fig. \ref{fgr:sketch_hysteresis},
for a droplet to advance on a liquid infused surface, the oil-water 
contact line has to recede and the oil-gas contact line has to advance. Similarly, a receding droplet requires the oil-water contact 
line to advance and the oil-gas contact line to recede; (iii) For $\theta^{\rm CB}_{\rm ow} > 0^\circ$ and $\theta^{\rm CB}_{\rm og} = 0^\circ$,
contact line pinning only occurs at the oil-water contact line, while (iv) for $\theta^{\rm CB}_{\rm ow} = 0^\circ$ and $\theta^{\rm CB}_{\rm og} > 0^\circ$,
pinning only takes place for the oil-gas contact line.

In our model the contact angles are defined with respect to the oil phase, and are required 
to be small to guarantee the validity of the hemi-wicking criterion, Eq. \eqref{eq:wicking}. The complementary angles,
defined with respect to the water and gas phases are therefore large, and the analogy to superhydrohobic materials
is appropriate. 

A large body of work on contact angle hysteresis on superhydrophobic materials leads to the
surprisingly simple result that the liquid (e.g. water) advancing contact angle occurs for 
$\theta^{\rm A}_{\rm wg}=180^\circ$ \cite{Kusumaatmaja2007,Schellenberger2016},
where deviations reported in literature are most likely due to experimental difficulties 
of measuring very large angles. The estimate for the receding angle is more
debated, and several models have been proposed in the literature. These include 
(i) the sparse defect model proposed by Joanny and DeGennes \cite{Joanny1984}, and experimentally tested by 
Reyssat and Quere \cite{Reyssat2009}, which suggests the receding contact angle 
has a logarithmic dependence with respect to the pillar spacing; 
(ii) thermodynamic approaches based on a linear average of the contact angle along the contact line\cite{Raj2012}, and 
(iii) the Cassie-Baxter model based on the area average\cite{McHale2004a}. 
It is not in the scope of this work to assess the accuracy of such models in general, but we remark that 
the thermodynamic Cassie-Baxter approach is more aligned with the approximations assumed here.
The sparse defect model is not consistent with the requirement of dense patterns, 
while the linear averaging model implies a strong effect of the orientation of the contact line on the global 
shape of a drop, which appears negligible in the currently available experimental data \cite{Guan2015,Schellenberger2016}. 

\begin{figure}[tb]
\centering
  \includegraphics[width=0.7\columnwidth]{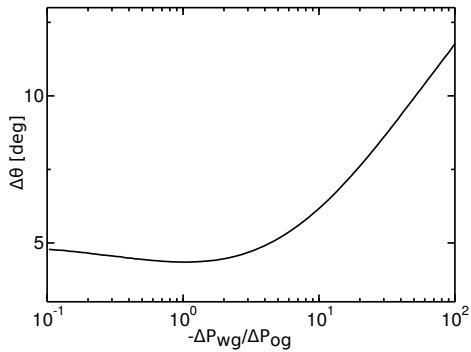}
  \caption{Contact angle hysteresis ($\Delta \theta$) of a water droplet on a liquid infused surface with 
  $\theta_{\rm o}=30^\circ$, $\theta_{\rm w}=\theta_{\rm g}=165^\circ$,
   $\theta^{\rm CB}_{\rm ow}=30^\circ$ and $\theta^{\rm CB}_{\rm og}=15^\circ$.
  For this set of parameters, the two sets of data shown in Fig. \ref{fgr:pressure}(a) in fact correspond to the 
   advancing ($\theta^{\rm R}_{\rm ow}=0^\circ$, $\theta^{\rm A}_{\rm og}=15^\circ$)
  and receding ($\theta^{\rm A}_{\rm ow}=30^\circ$, $\theta^{\rm R}_{\rm og}=0^\circ$) contact angles.
  We only focus on the definition of the apparent contact angle as defined at the triple junction, $\theta_{\rm app}$.
   }
  \label{fgr:hysteresis}
\end{figure}

\begin{figure*}[tb]
\centering
  \includegraphics[width=1.5\columnwidth]{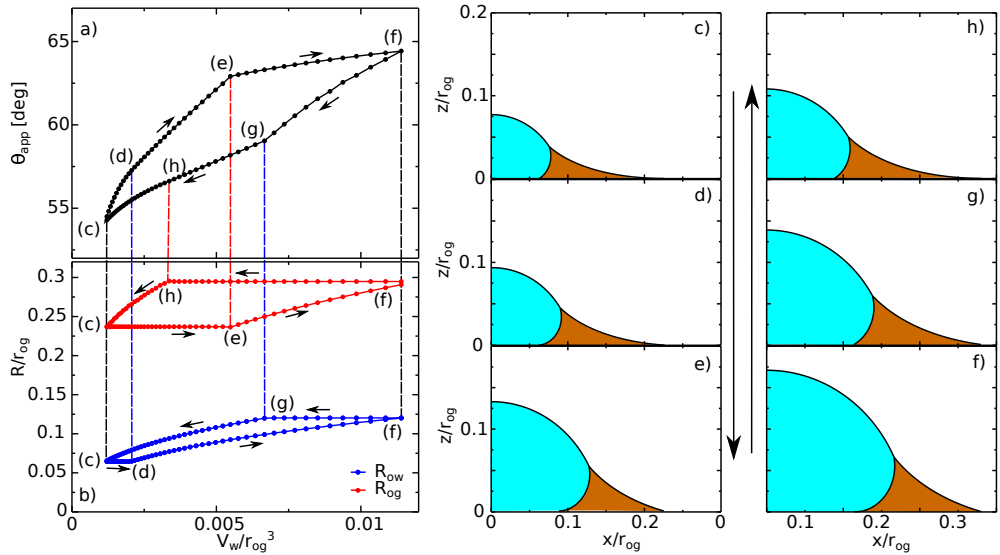}
  \caption{A typical contact angle hysteresis loop for a water droplet on a liquid infused surface. 
  Here $\theta_{\rm o}=30^\circ$, $\theta_{\rm w}=\theta_{\rm g}=165^\circ$,
   $\theta^{\rm CB}_{\rm ow}=30^\circ$ and $\theta^{\rm CB}_{\rm og}=15^\circ$. 
   Panel (a) shows the apparent contact angle of the droplet as a function of its volume,
   while panel (b) shows the radii of the oil-water ($R_{\rm ow}$) and oil-gas ($R_{\rm og}$) contact lines.
   The oil pressure, $\Delta P_{\rm og} = -2\gamma_{\rm og}/r_{\rm og}$,  is kept constant in these calculations, 
   where  $r_{\rm og}$ is the radius of curvature of the oil-gas interface. We use $r_{\rm og}$ 
   to normalise the droplet volume and the contact line radii. (c-h) Drop morphologies as indicated in panels (a) and (b).   
  }
  \label{fgr:hysteresis_loop}
\end{figure*}

Contact angle hysteresis is usually evaluated employing two alternative experimental approaches. 
The first one relies on applying a body force to the droplet\cite{Semprebon2014}, and the
advancing and receding angles are measured at the front and back of the droplet
just before it starts to move. To aid the discussion, let us now consider a specific example 
where the oil-water and oil-gas (Cassie-Baxter) contact angles are respectively $\theta^{\rm CB}_{\rm ow}=30^\circ$
and $\theta^{\rm CB}_{\rm og}=15^\circ$. The Neumann angles are chosen to be
$\theta_{\rm o}=30^\circ$ and $\theta_{\rm w}=\theta_{\rm g}=165^\circ$.
Based on our discussion in the previous paragraph, the conditions for an advancing contact line are
$\theta^{\rm R}_{\rm ow}=0^\circ$ and $\theta^{\rm A}_{\rm og}=\theta^{\rm CB}_{\rm og}=15^\circ$, while for a receding 
contact line we have $\theta^{\rm A}_{\rm ow}=\theta^{\rm CB}_{\rm ow}=30^\circ$ and $\theta^{\rm R}_{\rm og}=0^\circ$.
As such, the curves in Fig. \ref{fgr:pressure} represent the advancing and receding apparent contact angles for
a water drop on a liquid infused substrate with the aforementioned Cassie-Baxter contact angles, parametrized by the pressure
ratio $-\Delta P_{\rm wg}/ \Delta P_{\rm og}$. The contact angle hysteresis is defined as the 
difference between the advancing and receding contact angles, 
$\Delta \theta_{\rm app} = \theta^{\rm A}_{\rm app} - \theta^{\rm R}_{\rm app}$, and it is shown in Fig. \ref{fgr:hysteresis}
as a function of $-\Delta P_{\rm wg}/ \Delta P_{\rm og}$. The contact angle hysteresis shows a strong dependence on the pressure ratio 
(or equivalently the size of the oil ridge relative to the water droplet). It increases logarithmically in the
limit of $ -\Delta P_{\rm wg}/ \Delta P_{\rm og} \rightarrow \infty$, and it approaches a constant value 
as $-\Delta P_{\rm wg}/ \Delta P_{\rm og} \rightarrow 0$. Interestingly, the curve is also non
monotonic, and exhibits a shallow minimum close to $-\Delta P_{\rm wg}/ \Delta P_{\rm og} =0.2$.
This is in contrast with binary systems (e.g. water-gas on a solid surface), where the advancing and receding 
angles (correspondingly, contact angle hysteresis) can be regarded as constant material parameters.

The analytical expressions in Eqs. \eqref{eq:angle2},  \eqref{eq:anglepresssol0} and  \eqref{eq:linetension}
can be modified to predict advancing and receding contact angles in the limit of small oil ridge,
with the following replacement: 
$\theta^{\rm R}_{\rm ow}=0^\circ$, 
$\theta^{\rm A}_{\rm ow}=\theta^{\rm CB}_{\rm ow}$,  
$\theta^{\rm A}_{\rm og}=\theta^{\rm CB}_{\rm og}$, 
$\theta^{\rm R}_{\rm og}=0^\circ$.
In the limit of $-\Delta P_{\rm wg}/ \Delta P_{\rm og} \rightarrow 0$, we obtain
\begin{equation}
\label{eq:angleAdv}
\cos\theta^{S,A}_{\rm app} = - \frac{\gamma_{\rm ow}}{\gamma_{\rm wg}} + \cos\theta^{\rm CB}_{\rm og} \frac{\gamma_{\rm og}}{\gamma_{\rm wg}},
\end{equation}
and 
\begin{equation}
\label{eq:angleRec}
\cos\theta^{S,R}_{\rm app} = - \cos\theta^{\rm CB}_{\rm ow} \frac{\gamma_{\rm ow}}{\gamma_{\rm wg}} + \frac{\gamma_{\rm og}}{\gamma_{\rm wg}}.
\end{equation}
Furthermore, the resisting force due to contact angle hysteresis is given by
\begin{equation}
\label{eq:force}
F = 2 R \gamma_{\rm wg} \Delta \cos\theta,
\end{equation}
where $R$ is the contact radius and $\Delta \cos\theta = \cos\theta^{\rm R} - \cos\theta^{\rm A}$. 
We can straightforwardly obtain the expression for
$\Delta \cos\theta^{S}_{\rm app}=\cos\theta^{S,A}_{\rm app}-\cos\theta^{S,R}_{\rm app}$
by combining Eqs. \eqref{eq:angleAdv} and \eqref{eq:angleRec}. Here we assume the action of the
body force does not significantly deform the droplet. 
The resulting closed form expression once again
can be interpreted as a 'weighted sum' of the contact angle hysteresis for the oil--water and oil--gas contact lines.
Similar expressions for $\Delta \cos\theta_{\rm app}$ can also be obtained for small but finite $-\Delta P_{\rm wg}/ \Delta P_{\rm og}$ 
by exploiting Eqs. \eqref{eq:anglepresssol0} or  \eqref{eq:linetension}.

The second approach to measure contact angle hysteresis is by varying the volume of the water droplet \cite{Ruiz-Cabello2011}.
It is important to keep in mind that this protocol, unlike the previous one, involves measurements at
different pressure ratio $-\Delta P_{\rm wg}/ \Delta P_{\rm og}$. To elucidate the relevance of $-\Delta P_{\rm wg}/ \Delta P_{\rm og}$, 
we report a typical hysteresis loop in Fig. \ref{fgr:hysteresis_loop}.  As before, we consider 
$\theta^{\rm CB}_{\rm ow}=30^\circ$, $\theta^{\rm CB}_{\rm og}=15^\circ$, 
$\theta_{\rm o}=30^\circ$ and $\theta_{\rm w}=\theta_{\rm g}=165^\circ$ 
such that the data shown in Fig.\ref{fgr:pressure}(a) are the 
advancing and receding apparent contact angles as function of the pressure ratio
for this set of parameters.

Let us begin with the drop configuration shown in panel (c) of Fig. \ref{fgr:hysteresis_loop}. 
When the drop volume is increased, the apparent contact angle also increases. Here both the oil-water 
and oil-gas contact lines are pinned. At (d), the oil-water contact angle locally reaches 
$0^\circ$, and as a result, its contact line depins. With increasing the droplet 
volume, the oil-water contact line slides while the oil-gas one remains pinned.
From configuration (e), both contact lines reach their corresponding
advancing and receding contact angles, and become free to move. 
Correspondingly we observe a clear change of slope in 
the volume-contact angle relation. 
Once reaching (f), we reverse the process and decrease the droplet volume. 
Similar to the advancing scenario, initially both contact lines are pinned.  As before, the depinning of the 
oil--water and oil--gas contact lines are not simultaneous, 
and occur at (g) for the oil--water contact line and at (h) for the 
oil--gas contact line. Both contact lines move freely from (h) to (c), which 
is our starting configuration.

\section{Discussion}
\label{sec:Conclusions}

In this work we have theoretically investigated the apparent contact angle and contact angle hysteresis 
of a droplet on liquid infused surfaces (LIS). We derived a closed form expression for the apparent contact angle
in the limit of vanishing oil ridge that captures the energy balance of the three fluid phases in contact with the solid substrate.
Moreover, we computed the first order correction to the contact angle accounting for the influence of a small but finite oil 
wetting ridge surrounding the droplet, and showed that the correction term can be interpreted as a negative line tension.
We also employed numerical calculations to explore the full range of negative oil-gas Laplace pressures, 
showing that the apparent contact angles indeed vary as a function of pressure. 
Unlike usual wetting scenarios involving two fluids (e.g. water--gas), the apparent contact angle for LIS cannot be regarded
as a constant material property. We further note that our analytical expressions are in excellent agreement with the numerical results.

By introducing appropriate models for pinning and depinning
of the oil--water and oil--gas contact lines, we showed how the analytical expression for the apparent contact angles
can be readily manipulated to predict contact angle hysteresis on liquid infused surfaces. 
We presented a typical contact angle hysteresis loop, where we demonstrated that
the depinning of the oil--water and oil--gas contact lines are in general not simultaneous.
Contact angle hysteresis on LIS also depends on the oil pressure, or alternatively
the relative size of the oil wetting ridge to the water droplet.
Numerical calculations indicate that the contact angle hysteresis is smaller
for large and negative oil pressure (small ridge), compared to small and negative oil pressure 
(large ridge). This finding provides a useful design principle for LIS, 
suggesting that the contact angle hysteresis can be tuned by the oil pressure, which
can be achieved for example by under-filling/over-filling the substrate with oil.

Our results so far are limited to equilibrium morphologies. A full characterisation of 
wetting dynamics on liquid infused surfaces is an important and open problem.
To this end, we recently developed a ternary free energy Lattice Boltzmann approach\cite{Semprebon2016},
well suited for handling the fluid dynamics of the water droplet and infusing oil, and
for taking into account the Neumann angles and wetting contact angles involved in the problem. 
Another important direction for future work is to investigate the possible presence of thin oil film
coating the surface corrugations and/or the water droplet \cite{Smith2013}, including the molecular mechanism
that determines the film thickness and its influence to the shape of the water droplet when
their length scales are comparable. When the infusing oil cloaks the droplet, the water-gas surface tension
is not the appropriate variable to use in Eqs. \eqref{eq:balance}, \eqref{eq:angle2} and  \eqref{eq:linetension}.
Instead, it should be replaced by a composite water-oil and oil water interfaces, 
$\gamma_{\rm wg} \rightarrow \gamma_{\rm ow}+\gamma_{\rm og} - \Delta(e)$, where the binding potential
$\Delta(e)$ is a function of the oil film thickness, and its form depends on the intermolecular interactions
of the fluids. We will explore this case in more details in future works. 

\section{Acknowledgements}
\label{sec:acknowledgements}

C.S. and H.K. are grateful to Dr. M. Brinkmann for inspiring comments on this manuscript, 
and thank Dr. M. Wagner, Dr. Y. Gizaw, Dr. P. Koenig, Prof. G. Mistura and M. S. Sadullah 
for fruitful discussions. We acknowledge Procter and Gamble (P\&G) for funding.

\bibliographystyle{apsrev4-1} 
\bibliography{references}

\end{document}